\documentclass[11pt]{article}
\usepackage[utf8]{inputenc}
\usepackage{amsfonts,url,epsfig,float,xcolor}
\usepackage{geometry} 
\usepackage{sectsty} 
\usepackage[normalem]{ulem}
\usepackage{booktabs,tabularx}
\usepackage{caption}

\sectionfont{\sffamily\bfseries\upshape\large}
\subsectionfont{\sffamily\bfseries\upshape\normalsize} 
\subsubsectionfont{\sffamily\mdseries\upshape\normalsize}
\makeatletter
\renewcommand\@seccntformat[1]{\csname the#1\endcsname.\quad}
\makeatother\renewcommand{\bibitem}{\vskip 2pt\par\hangindent\parindent\hskip-\parindent}

\makeatletter
\def\@maketitle{%
  \begin{center}%
  \let \footnote \thanks
    {\large \@title \par}%
    {\normalsize
      \begin{tabular}[t]{c}%
        \@author
      \end{tabular}\par}%
    {\small \@date}%
  \end{center}%
}
\makeatother

\newcommand{\btR}{\vspace{-.25in}\begin{quotation}\begin{small}\noindent\begin{verbatim}}

\title{\bf The ladder of abstraction in statistical graphics\footnote{We thank Howard Wainer, Ron Yurko, Dianne Cook, Alessandra Casella, and two anonymous reviewers for helpful comments.}\vspace{.1in}}

\author{Andrew Gelman and Kaiser Fung\vspace{.1in}}

\date{12 Sep 2025\vspace{-.1in}}

\begin{document}\sloppy
\maketitle

\begin{abstract}

Graphical forms such as scatterplots, line plots, and histograms are so familiar that it can be easy to forget how abstract they are.  As a result, we often produce graphs that are difficult to follow.  We propose a strategy for graphical communication by climbing a ladder of abstraction (a term from linguistics that we borrow from Hayakawa, 1939), starting with simple plots of special cases and then at each step embedding a graph into a more general framework.  We demonstrate with two examples, first graphing a set of equations related to a modeled trajectory and then graphing data from an analysis of income and voting.
\end{abstract}

\section{Motivation}

A primary goal of statistics is generalizing from data. Even the most basic statistical construct, the statistical average, is an abstract concept. Modern statistical models contain layers of abstraction.  We think such abstraction poses robust challenges to the effective communication of statistical findings.  While graphs are intended to facilitate understanding of data, they are often difficult to follow because they too are abstractions.

We propose a strategy for graphical communication by climbing a ladder of abstraction, a term from linguistics that we borrow from Hayakawa (1939) who explains how words exist at different levels of abstraction and proposes that an awareness of such structure enables meaningful language.  His framework has proven practical and popular, initially among education researchers adapting it as a teaching tool for narrative writing (e.g., Moffett, 1968).  More broadly, social scientists use Hayakawa's semantic principle to bring structure to text treated as data (e.g., Hurst and Walker, 1972).  Later applications appeared in more analytical fields, such as accounting (Guthrie, 1972), safety management (Metzgar, 1978), cartography (Jahnke et al., 2011), and program evaluation (Leviton, 2015).

In a section about defining words, Hayakawa argued that proper definitions point down the ladder of abstraction.  This accords with our view that an effective graphical communication strategy should start from simple plots of special cases and then, in calculated steps, introduce layers of abstraction.  We will demonstrate this strategy using two examples.

In this article we limit our discussion to scatterplots, line plots, and grids of plots with quantitative axes on the Cartesian plane.  We imagine that similar principles should apply to other graphical forms, both static and dynamic, in which complexity can be built up, step by step.

\section{Visualizing a mathematical model:  basketball shooting}

Consider the following simple problem in applied mathematics:  What is the optimal angle for shooting a basketball, if you would like to be able to throw the ball as softly as possible and still reach the hoop?  The solution will be a function of your distance from the hoop and the altitude at which you release the ball.

\begin{figure}
\vspace{-.9in}
\centerline{\includegraphics[width=.475\textwidth]{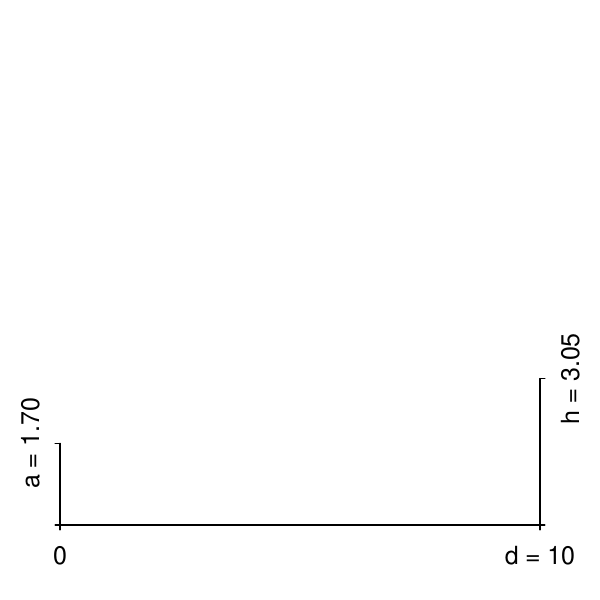}}
\vspace{-.1in}
\caption{\em Diagram of a person standing 10 meters away from a basketball hoop, with the ball released at an altitude of 1.7 meters and the hoop being 3.05 meters high.  This is the first step of a series of graphs that will culminate in displaying a optimal shooting angle.}\label{basketball_1}
\end{figure}

\begin{figure}
\vspace{-.9in}
\centerline{\includegraphics[width=.475\textwidth]{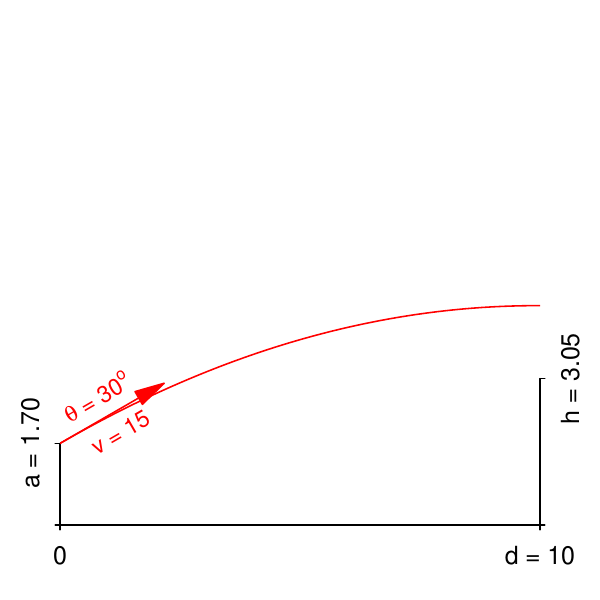}\hspace{.05\textwidth}\includegraphics[width=.475\textwidth]{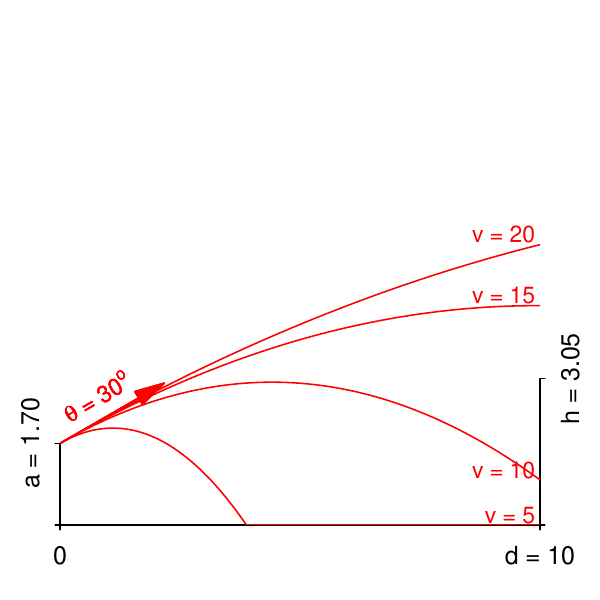}}
\vspace{-.1in}
\caption{\em (Left:) Trajectory of a shot taken at a $30^{\circ}$ angle at a velocity of 15 m/sec, as computed using Newtonian mechanics ignoring air resistance.  The shot goes too high and misses the hoop.  (Right:) Trajectories of a series of shots with different initial velocities.  The velocity required to reach the hoop must be somewhere between 10 and 15 m/sec. The left graph is at a low level of abstraction in that the red line directly shows the path of the ball.  The graph on the right is more abstract in that it shows several trajectories at once.}\label{basketball_2}
\end{figure}

We would like to reach a general solution but we start with the specific case in Figure \ref{basketball_1}:  a player standing $d=10$ meters from the basketball hoop, releasing the ball at an altitude of $a=1.7$ meters.  The hoop itself is $h=3.05$ meters off the ground.

The left plot in Figure \ref{basketball_2} shows the path of a ball shot at a $\theta=30^{\circ}$ angle at an initial velocity of $v=15 \mbox{m}/\mbox{sec}$.  From Newton's laws of motion and assuming no air resistance, the trajectory is $x(t) = v\cos(\theta)\,t$, $y(t) = a + v\sin(\theta)\,t - 0.5gt^2$, where $g=9.8 \mbox{m}/\mbox{sec}^2$ is the gravitational constant.  The red line on this graph directly shows the ball's trajectory.  The only way to make it more vivid and less abstract would be an animated presentation that dynamically shows the moving ball.

We then ask what would happen at this same launch angle but with different initial speeds.  The right plot in Figure \ref{basketball_2} shows trajectories corresponding to $v=5$, 10, 15, and 20 m/sec.  This graph is slightly more abstract than the one on the left in that it expands the coverage from a single case to a range of possible paths. In addition, because all the curves start at the same point, we moved the label of the initial velocity at the end of each path, which is not quite intuitive.  This is an example of the conceptual distancing that can be required when increasing the information content of a plot.  Adding lines to a plot does not necessarily make it more abstract, but in this case the additional lines represent another dimension in the analysis and another possible direction of comparison. We find it helpful to present both graphs of Figure \ref{basketball_2} in order to guide the reader through the process of abstraction.

Our next step is to figure out the initial velocity needed to reach the basket, given the shooting angle.  This is a straightforward algebra problem.  First we figure out the time $T$ required for the ball to reach the horizontal position of the hoop:  $d = v\cos(\theta)\,T$, thus $T=\frac{d}{v\cos(\theta)}$.  Then we solve for the initial velocity $v$ so that the ball's vertical postion at time $T$ is that of the hoop:  $h = a + v\sin(\theta)\frac{d}{v\cos(\theta)} - 0.5 g\frac{d^2}{v^2(\cos(\theta))^2} = a + d\tan(\theta) - 0.5\frac{gd^2}{v^2(\cos(\theta))^2}\,$, so
\begin{equation}\label{v}
v=\sqrt{\frac{0.5gd^2}{(\cos(\theta))^2(d\tan(\theta) + a - h)}}\, .
\end{equation}
Plugging $a=1.70, \theta=30^{\circ}, d=10, h=3.05, g=9.8$ into this equation yields $v=12.2$.  The resulting trajectory is the blue line in Figure \ref{basketball_3}.  It is helpful to see it in the context of the red curves that show other velocities.

\begin{figure}
\vspace{-.9in}
\centerline{\includegraphics[width=.475\textwidth]{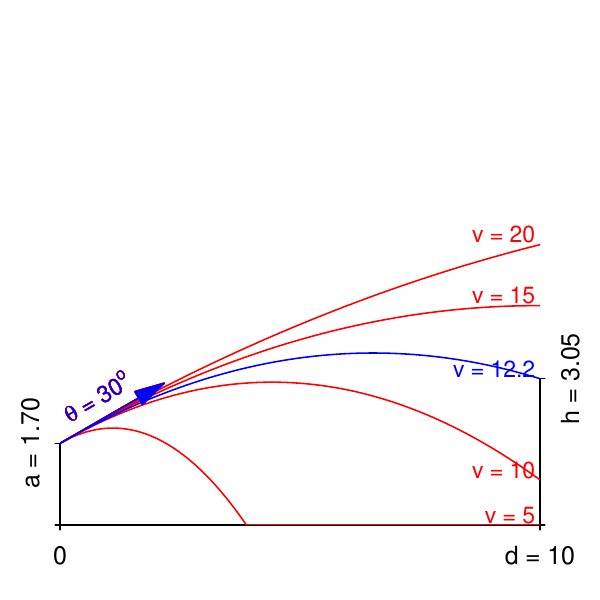}}
\vspace{-.1in}
\caption{\em A shot at angle $30^{\circ}$ from altitude 1.7 m, from a distance of 10 m, will reach the hoop if its initial velocity is 12.2 m/sec.}\label{basketball_3}
\end{figure}

\begin{figure}
\centerline{\includegraphics[width=.475\textwidth]{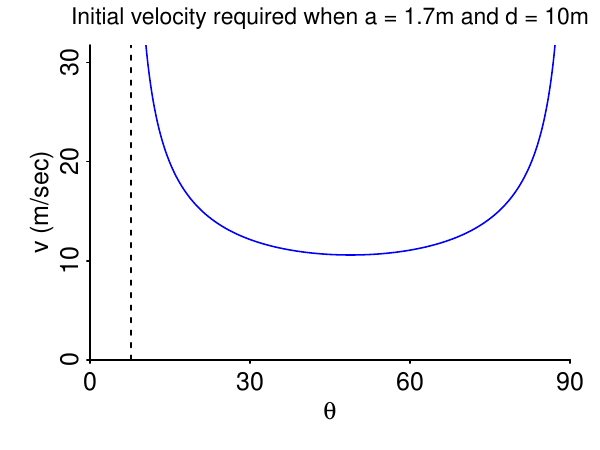}\hspace{.05\textwidth}\includegraphics[width=.475\textwidth]{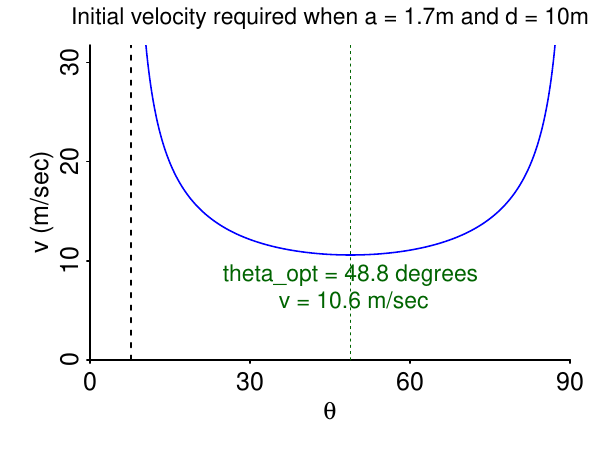}}
\vspace{-.1in}
\caption{\em (Left:)  Velocity required to reach the basketball hoop for a shot released at an altitude of 1.7 meters, 10 meters away from the basket. (Right:)  On this graph, the required velocity $v$ is minimized at an angle $\theta=48.8^{\circ}$, at which point $v = 10.6$ m/sec.  The blue color of the lines matches the blue used to display the hoop-reaching solution in Figure \ref{basketball_3}.}\label{basketball_45}
\end{figure}

\begin{figure} \centerline{\includegraphics[width=.475\textwidth]{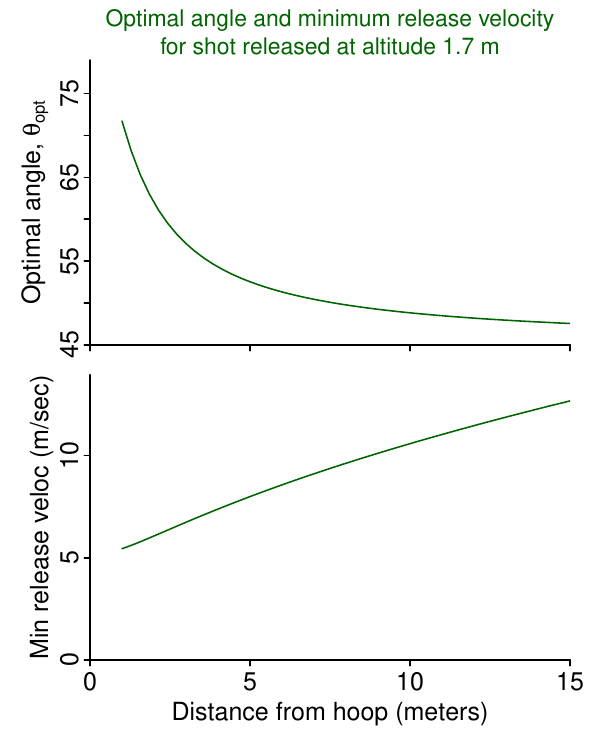}\hspace{.05\textwidth}\includegraphics[width=.475\textwidth]{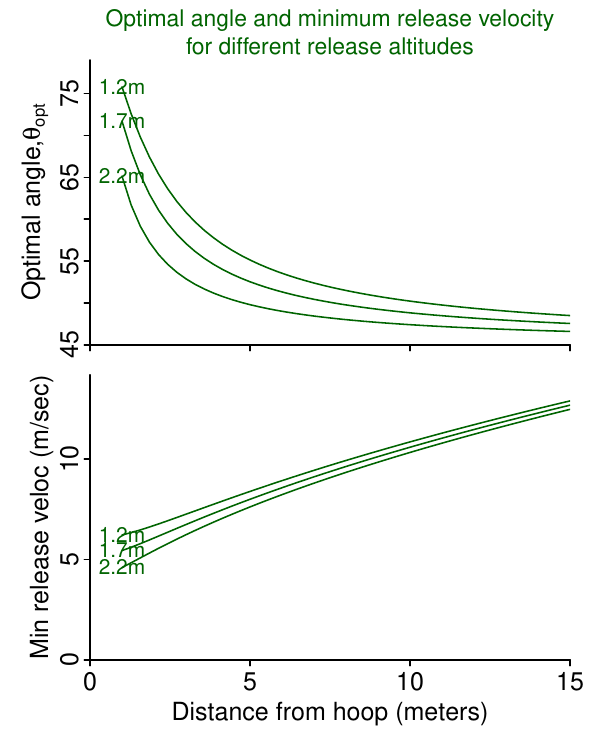}}
\vspace{-.1in}
\caption{\em (Left:)  Shooting angle requiring the softest shot and the corresponding release velocity, as a function of the distance from the hoop, for a shooter releasing the ball at an altitude of 1.7 meters.  (Right:)  Same graphs, also showing the solutions for release altitudes of 1.2 and 2.2 meters.  The green color of the lines matches the green used to display the minimum-initial-velocity solution in Figure \ref{basketball_45}.}\label{basketball_67}
\end{figure}

Now that we have an equation for the required initial velocity for the ball to reach the hoop, we can plot it as a function of the shooting angle, $\theta$.  The left plot of Figure \ref{basketball_45} shows the result.  For the lowest angles ($\theta\leq \tan^{-1}(\frac{h-a}{d})$, indicated by the dashed vertical line on the plot), the ball will go below the hoop no matter how hard it is thrown.  The lowest initial velocities are needed for shots taken from between $30^{\circ}$ and $60^{\circ}$, with harder shots required when the angle goes outside that range.

We can then use a general-purpose numerical optimizer to find the value of $\theta$ that minimizes (\ref{v}), given $a=1.70,d=10,h=3.05,g=9.8$.  The result is $\theta=48.8^{\circ}$, yielding $v=10.6$.  This result is displayed as a green dotted line in the right plot of Figure \ref{basketball_45}.  Again, we find it helpful to display both graphs, as they show the development of the analysis:  first the curve, then the minimum, which we label as the ``optimum'' in the sense of requiring the softest throw to reach the hoop.

Compared to Figure \ref{basketball_3}, Figure \ref{basketball_45} represents a leap in abstraction:  the curve is not a trajectory, and it embodies the solution to an equation.  By seeing both charts, readers can observe the switch from physical space to an abstract space: the blue line in Figure \ref{basketball_3} is the single dot at $\theta=30^{\circ}$ in Figure \ref{basketball_45}, in which the angle $\theta$ and the initial velocity $v$ are now represented by linear coordinates.  Again, in statistics we are so familiar with this sort of graph that we don't always remember how abstract it is.

The punch line of Figure \ref{basketball_45} is the recommended shooting angle, $\theta_{\rm opt}$, and corresponding initial velocity $v$, for a shot taken at an altitude of 1.7 meters, 10 meters from the basket.  Naturally, we should ask how these parameters change as a function of distance from the basket.  Obviously $v$ will increase with distance, but at what rate?  And what will happen to $\theta_{\rm opt}$?

We answer this question by unfixing $d=10$ and instead considering a grid of values of $d$, ranging from 1 to 15 meters from the hoop.  For each value of $d$, we numerically optimize and find the value of $\theta$ that minimizes (\ref{v}), along with the corresponding value of $v$.

The left panel of Figure \ref{basketball_67} shows the results.  When you are standing close to the basket you need to shoot at a steep angle to reach the hoop; as the distance increases, the optimal angle approaches $45^{\circ}$.  The required release velocity increases gradually, at a rate slightly less than linearly with distance.  We have lined up the two plots vertically so they can share an $x$-axis and the reader can see how both $\theta_{\rm opt}$ and $v$ vary together. 

Finally, we gain one more level of generality by considering different release points:  1.2 m, 1.7 m, and 2.2 m off the ground, which might correspond to shooting from chest level, head level, or above the head.  The right panel of Figure \ref{basketball_67} shows curves for each of these three heights.  Unsurprisingly, releasing from a greater altitude requires a lower shooting angle and a lower initial velocity.

Readers experience a similar step up in abstraction going from Figure \ref{basketball_45} to Figure \ref{basketball_67} as they did from Figure \ref{basketball_3} to Figure \ref{basketball_45}.  The curve in Figure \ref{basketball_45} is now one pair of dots in the paired charts of Figure \ref{basketball_67}.  The coupling of the angle and speed parameters also carries over from Figure \ref{basketball_45}.

The right panel of Figure \ref{basketball_67} represents the greatest level of abstraction in this series of graphs, and, again, we find it helpful to build up to it rather than simply presenting these results and then expecting readers to follow.

We have also used graphical cues as much as possible to make connections between the graphs in the series.  When two graphs share the same graphical space, as defined by the variables in the axes, the axis ranges and aspect ratios are kept constant.  When the space definition shifts, we include sufficient information to convey the relationships between the two spaces.

In addition, throughout this example we have attempted to guide the reader using a succession of colors:
\begin{itemize}
\item {\em Black} for the basketball court in Figures \ref{basketball_1}--\ref{basketball_3} and the axes in Figures \ref{basketball_45}--\ref{basketball_67},
\item \textcolor{red}{\em Red} lines for the trajectories in Figures \ref{basketball_1}--\ref{basketball_3}.
\item \textcolor{blue}{\em Blue} line for the trajectory in Figure \ref{basketball_3} that reaches the hoop, and also for the curves in Figure \ref{basketball_45} that represent the initial velocity as function of angle for these shots,
\item \textcolor{green}{\em Green} for the optimal angle that minimizes the required initial velocity in the right plot of Figure \ref{basketball_45}, and also for the curves in Figure \ref{basketball_67} that show this optimal angle and corresponding initial velocity as a function of distance and release altitude.
\end{itemize}
The colors climb the ladder of abstraction, from direct mapping of trajectories to the solutions of a series of equations.  When presenting the series of graphs we do not draw attention to the colors---that might feel like overkill and distract readers from the content being graphed---but we hope the colors help, in the same way that good design can make a tool more effective even for users who are unaware of these choices.

\section{Visualizing data:  income and voting within and between states}

In a book on quantitative trends in politics by one of the co-authors (Gelman et al., 2009), we were particularly attuned to the challenges of communicating to a general audience.  It was important to us to use graphs and not just text because we wanted to actively engage readers in the process of understanding and discovery.  Rather than simply {\em saying}, for example, that richer people were more likely than poorer people to vote for Republicans, but Democrats did better in rich states, we wanted to {\em show} it.

Figure \ref{redblue_1} juxtaposes the two patterns for George W. Bush in 2004, showing he did better among richer voters but worse within poorer states.  Despite using standard chart forms, both these graphs are abstract, but in different ways: in the left plot, each dot represents the abstract construct of an income subgroup, a set of survey respondents with a common response to the income question; in the right plot, the horizontal and vertical axes establish an abstract space, in which each state is represented by a point, given by its average income and vote share.

The contrasting patterns for individuals and states motivated us to look at income and voting within states.  Again, we climb the ladder of abstraction in Figure \ref{redblue_2}, by choosing three states (Mississippi, Ohio and Connecticut) to represent poor, middle-income, and rich states respectively.  Without such generalization, Figure \ref{redblue_2} would contain 50 lines and would be close to unreadable.

While the axis labels suggest income subgroups, you might notice the smooth lines rather than the connected points in the left panel of Figure \ref{redblue_1}.  In Figure \ref{redblue_2}, the lines display summaries of a fitted hierarchical logistic regression model---each of the lines is a part of a logistic curve, with each being on a narrow enough range that the lines appear nearly straight.  It might seem like cheating in Figure \ref{redblue_2}  to display fitted curves rather than raw data, but you can think of these as smoothed data or as an alternative to presenting logistic regression coefficients.

\begin{figure} \centerline{\includegraphics[height=.38\textwidth]{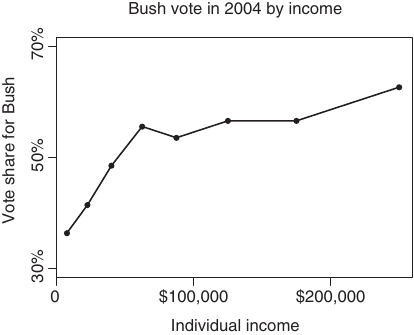}\hspace{.1\textwidth}\includegraphics[height=.38\textwidth]{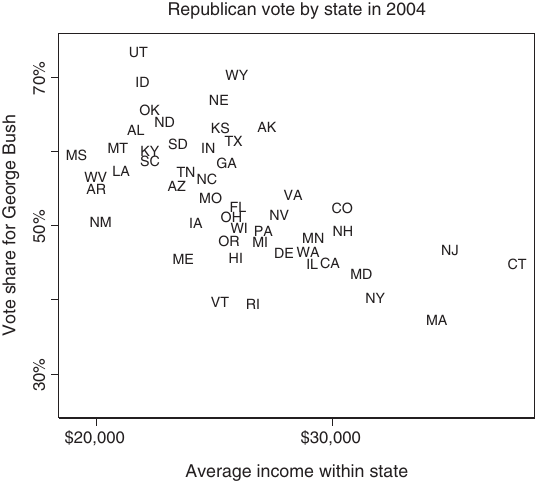}}
\caption{\em (Left:)  From exit polls, national voting patterns showing that George W. Bush,the Republican candidate for president in 2004, did better among richer voters. (Right:)  At the same time, Bush was more successful in poorer states.}\label{redblue_1}
\end{figure}

\begin{figure} \centerline{\includegraphics[width=.5\textwidth]{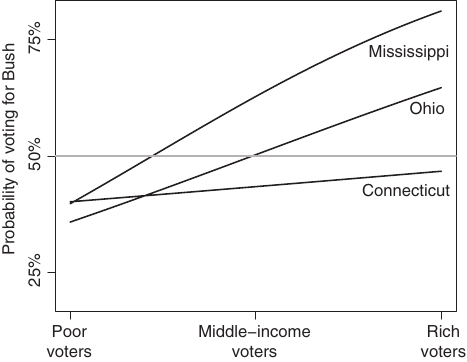}}
\caption{\em From exit polls, George W. Bush's vote share in 2004, as a function of income, in a poor state, a middle-income state, and a rich state.}\label{redblue_2}
\end{figure}

While Figure \ref{redblue_2} draws out the income and voting patterns within state subgroups, in Figure \ref{redblue_3} we add another feature to illustrate the between-state pattern, finally achieving our goal of juxtaposing divergent patterns within and between states.  On top of each state's line, open circles locate the five income subgroups and their respective probabilities of voting Republican, with the sizes of the circles proportional to the number of voters in each category, relative to the income distribution of the electorate as a whole, while the solid circle plots the Republican candidate's vote share in the state against the average income in the state on this five-point scale.  The startling insight from Figure \ref{redblue_3} is the opposite directions of correlation depending on whether one looks between or within states.  The pattern of the three solid circles in each graph shows how poorer states are more Republican-leaning, even while the open circles for each line indicates a positive correlation between income and voting Republican within each state.  To draw another connection between graphs, the lower-right plot in Figure \ref{redblue_3} corresponds to 2004, the same year as displayed in Figure \ref{redblue_2}, but the lines are slightly different because the estimates come from a different data source.

\begin{figure} \centerline{\includegraphics[width=.7\textwidth]{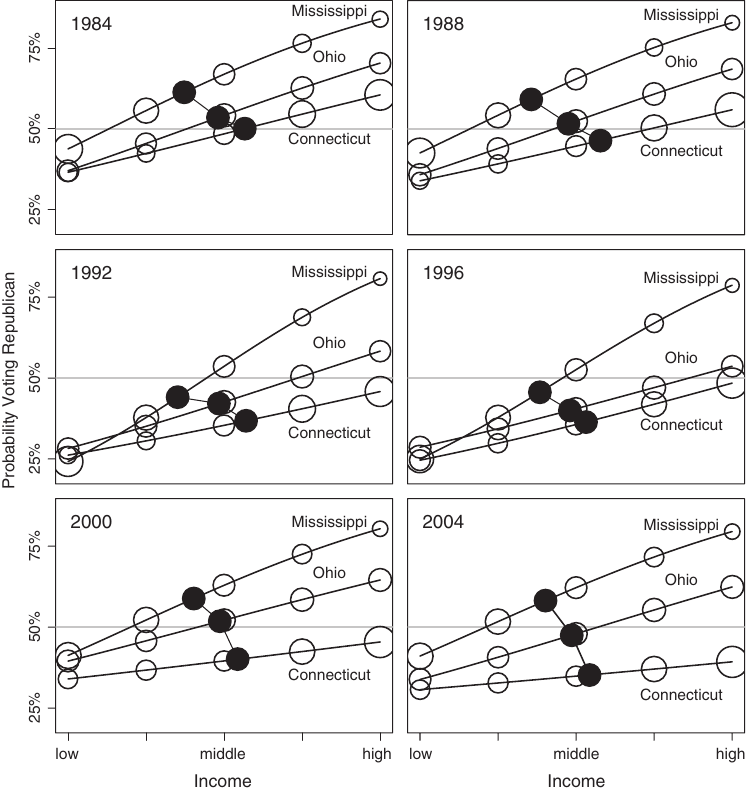}}
\caption{{\em Probability of Republican vote as a function of income in poor, middle-income, and rich states in six straight presidential elections. The open circles on the plot show the relative size of each income group in each state (thus, Mississippi has more poor people than average, and Connecticut has more rich people). The solid circles show the average Republican vote and average income in each state.}}\label{redblue_3}
\end{figure}

We considered options to include uncertainty estimates in these graphs, such as error bars or shaded uncertainty zones, but decided not to, as we were worried about cluttering that would make it more difficult to make visual comparisons.  In addition, the variation of the estimates from year to year gives some sense of estimation uncertainty.  In any case, there is no right answer here; there is an inevitable tradeoff between clarity and the presentation of more information.

As it stands, Figure \ref{redblue_3} is a leap of abstraction compared to Figure \ref{redblue_2}.  Each state is further represented by its average income, rather than continuous or categorical measures of income.  By expanding from one to six presidential election cycles, we have to generalize from voting for Bush to voting for the Republican candidate.

Indeed, one could take another step of abstraction and summarize each fitted curve by an intercept and slope on the logistic scale (with the income predictor coded as $-2,-1,0,1,2)$, so the intercept would be directly interpretable and then make a scatterplot of slope vs.\ intercept with two-letter state abbreviations.  Such a graph could be difficult to explain on its own but could be a natural followup to Figure \ref{redblue_2}, showing fifty states instead of just three.

\begin{figure} \centerline{\includegraphics[width=.6\textwidth]{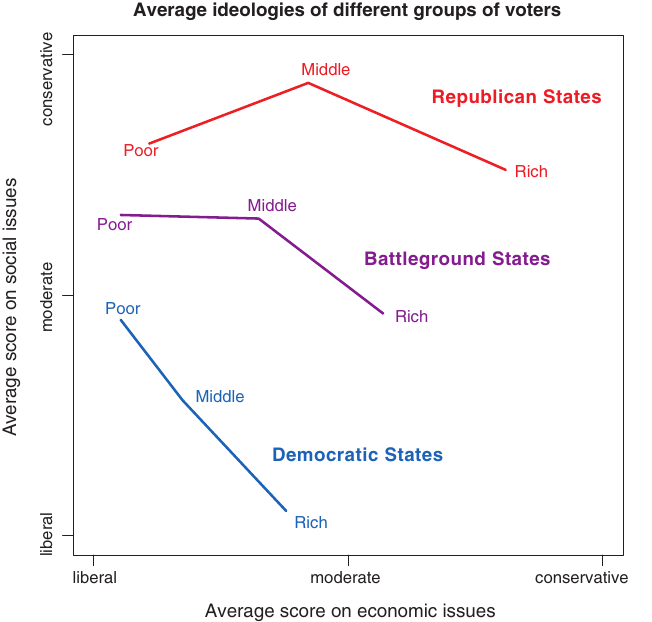}}
\caption{\em From a large national survey in 2000, average positions of respondents on social and economic issues among voters in different income levels, within Republican-leaning, battleground, and Democratic-leaning states.  Richer voters are consistently more conservative on economic issues, but the patterns on social issues are more complicated.}\label{redblue_4}
\end{figure}

To better understand what was happening with income and voting, we went to polling data and created left-right scores on economic and social issues by combining several survey items in each area.  The estimates were based on a large national poll from 2000 with enough data to get stable estimates for subgroups characterized by individual incomes and state political leanings.  Figure \ref{redblue_4} shows the result, which uses lines, colors, and labels to enable comparisons across voters and states in both dimensions (social and economic).

Figure \ref{redblue_4} scales the ladder of abstraction in several ways.  The economic and social scores represent aggregated responses to multiple survey questions while the income subgroups are determined by the income question alone.  These scores create an abstract space of ideology, in which individuals can be placed.  Instead of individuals, we plot income-state subgroups, which is another abstraction.  States are categorized as Republican-leaning, battleground, or Democratic-leaning.

\begin{figure} \centerline{\includegraphics[width=.8\textwidth]{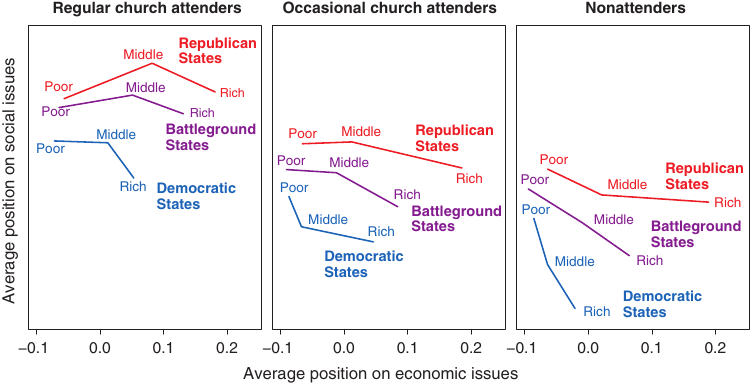}}
\caption{\em Continuation of Figure \ref{redblue_4}, further characterizing people by religious attendance.  This plot is complicated, but it is possible to read it by building up from earlier, less-abstract graphs.}\label{redblue_5}
\end{figure}

Once Figure \ref{redblue_4} has been understood, we can go one more level of abstraction to Figure \ref{redblue_5}, which displays the same comparisons but separately for people who attend church regularly, occasionally, or not at all.  We did not want to clutter the display with standard errors for all these averages, but uncertainty can be deduced from apparently random variation in the plots.

Figure \ref{redblue_5} represents the greatest level of abstraction in this series of graphs.  Presented on its own, it would take some effort to understand well, which is why we find it helpful to lay out the path in several steps, starting from raw survey data on income and voting and gradually adding more information.  In designing the graphs, we have used certain strategies.  Specific cases are frequently presented first, then generalized individuals, and finally subgroups that are abstractly defined by models.  In statistics, we are so used to talking about averages, without recognizing that they represent an abstract construct.

The effectiveness of graphs such as Figure \ref{redblue_5} will depend on the problem being studied and on the data being plotted.  Here it helped that the lines for the three groups of states were clearly separated and that the paths from poor to middle-income to rich voters had generally similar trajectories.  The cleaner the data, the less burden is placed on statistical modeling.  As with any applied problem, this is not the end of the story.  There are many other ways of looking at data on income and voting, and the patterns we showed above have changed in the years since those graphs were published (see, for example, Feller et al., 2012, and Gelman and Azari, 2017).

\section{Discussion}

\begin{figure} \centerline{\includegraphics[width=.8\textwidth]{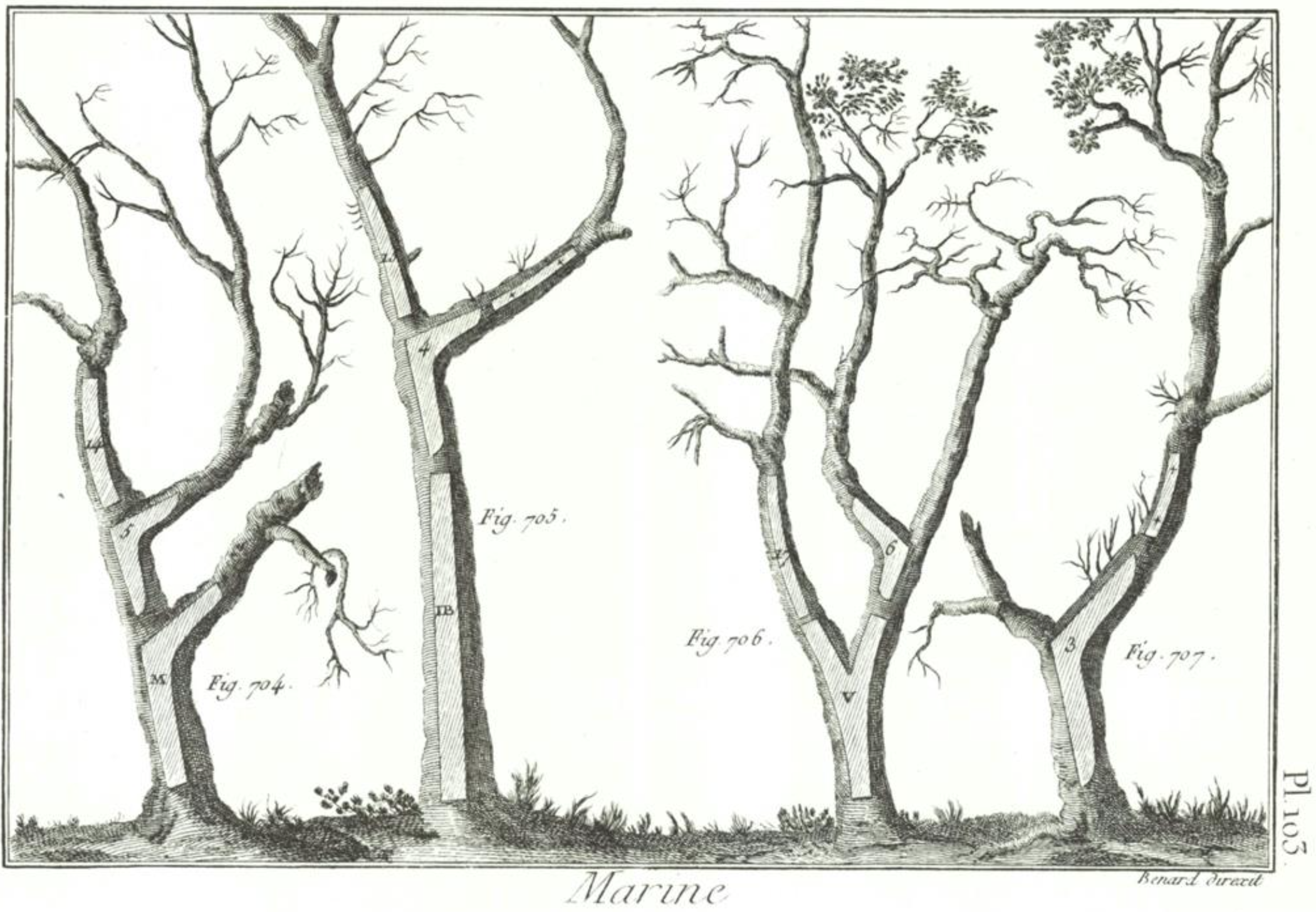}}
\caption{\em Graph from Panckoucke (1783) illustrating the portions of trees that could be used to construct different parts of a ship.  This is an example of a non-statistical visual representation at a low level of abstraction.} \label{trees}
\end{figure}

There is a long history of abstraction in visual representation, going far beyond statistical graphics.  Consider, for example, the phrase ``the map is not the territory'' (Korzybski, 1933) and the satirical vignette of Borges (1946) pointing out the uselessness of a perfectly faithful 1:1 scale map.  Figure \ref{trees} shows an amusing example discussed by Wainer and Friendly (2021) of an early infographic that cleverly conveys information directly through a non-abstract image.  Literature in cartography (Fabrikant, 2004) and information visualization (Cui et al., 2006, Viola and Isenberg, 2018) have considered the cognitive and perceptual aspects of abstraction in visual communication.  But when it comes to displaying mathematical or statistical relationships, we typically take for granted the abstractions of scatterplots, time series graphs, bar charts, and so on.

Our contribution in this paper is to demonstrate the general strategy of presenting an increasingly more abstract world using a series of graphs starting with simple plots of special cases.  This approach should be increasingly useful to statisticians and graphic designers in a world of big data and information visualization.  Statistical graphics can be beautiful, informative, and sophisticated, but they do not always speak for themselves.  It is well known that a graph can be much improved with a clear title and caption:  as the saying goes, a picture plus a thousand words is worth more than two pictures or two thousand words.  Here we have shown how it can also be helpful to build up to an informative graphic by explicitly showing intermediate graphs at lower levels of abstraction.

\newcommand{\thd}[1]{\small\bfseries #1}
\begin{table}
\centering
\setlength{\tabcolsep}{6pt}
\resizebox{0.9\textwidth}{!}{%
{\small
\begin{tabular}{@{} l l l @{}}
\thd{Strategy} & \thd{Basketball example} & \thd{Voting example} \\
\midrule
Define an abstract space or transform it & Figures 4 and 5 & Figure 8 \\
Use modeled or optimized values          & Figure 4            & Figures 7 and 8 \\
Use statistical averages                  &                     & Figure 8 \\
Expand coverage from single year to multiple years &             & Figure 8 \\
Expand sampling from single to multiple    & Figures 2 and 5  & \\
Define subgroups or select representative instances &           & Figures 7 and 8 \\
Allow a fixed parameter to vary                           & Figures 4 and 5  &
\end{tabular}
}
}
\caption{\em Some strategies of abstraction and where they are used in the two examples of this paper.  Each strategy represents a way that a statistical analysis or summary can be generalized.}\label{strategies-examples}
\end{table}

Throughout the two examples, we apply abstraction in various ways when moving up the ladder.  In each step, the prior graphs prepare the readers to understand the additional abstraction in the next graph, similar to how Hayakawa defines a more abstract word in terms of less abstract words.  These strategies are summarized in Table \ref{strategies-examples}.  When preparing a series of graphical displays, it could also be helpful to include example questions that might be asked at each level of abstraction, such as ``What does this graph show directly?'',  ``What comparisons are being facilitated by this display,'' ``How does this generalize across different subsets of the data?'', ``What assumptions are implicit in the model?'', and, often most importantly, ``How can this plot be misinterpreted and what can be done to avoid such misreadings?''

Modern treatments of data visualization or the grammar of graphics consider graphs not as static drawn objects but as overlays of information, so that a useful plot is built by adding elements to a general structure (Wilkinson, 2005, Wickham, 2016).  The ladder of abstraction discussed in the present paper represents a different sort of forward progression:  in addition to adding information to a graph, we are using increasing levels of generality.  Our ideas follow the spirit of the grammer of graphics, however, in considering statistical graphics as a form of communication of quantitative information to the reader, with the specific graph being just an instantiation of this goal.  As emphasized by Tufte (1983) and Cleveland (1985), choices in data display can be understood with respect to the specific goals of communication. In present paper, we emphasize the challenge that useful visualizations can be highly abstract representations of the information being presented, hence our proposal of stepping readers through a series of graphs. Here we have demonstrated this idea in two different applications of statistical graphics:  continuous curves and discrete data.

Once presented, the ladder of abstraction seems natural, but we have rarely seen it in published work, including our own.  Our usual approach in graphical communication has been to develop the graphs we want to display and then explain them in words.  The examples here suggest that a more explicit unfolding could be useful, by analogy to the paradigm in statistical workflow of fitting increasingly complicated approximations, with each understood in comparison to what came before, an idea we have found helpful for modeling as well as computation (Gelman et al., 2020).

In general, the need for abstraction in statistical graphics arises from the difficulty of representing complicated data structures and functional forms in two linear dimensions.  The particular sorts of abstractions needed will depend on the form of the observed data and of latent dimensions of interest, and the larger goals of the problem (finding an optimal shooting angle in our basketball example and exploring voting patterns as a function of income, religious attendance, and state in our political science example).  As statisticians we can start from these goals and, from there, figure out useful graphical displays, and then when presenting such displays we can recognize their abstract nature and present them by building up from simpler forms.

We have not offered any general strategy to construct such a series of plots, but we think the two examples in this paper demonstrate how it can be done in two very different situations.  We have deliberately chosen examples that differ in their general structure (mathematical functions vs.\ statistical data), subject area (physics vs.\ political science), and goals (optimization vs.\ exploration).  Even with all these differences, both problems benefit from starting at the core with directly-interpretable graphs and then gradually adding layers of complexity, as indicated in Table \ref{strategies-examples}.  We leave it to future researchers to further formalize these ideas and implement them into general-use software.

The present article has three messages.  First, some level of abstraction exists in all but the simplest graphs, and we should be aware of this when using graphs to communicate, especially to readers who are not familiar with data graphics.  Second, graphical abstraction is not just a difficulty; it also is a useful tool that allows us to understand patterns at a greater level of sophistication, and it can be valuable to see this by adding one level of abstraction at a time rather than trying to jump to a single plot that shows everything.  Finally, these principles apply not just to graphs of data but also to graphs of mathematical functions such as fitted models. 

\section*{References}

\noindent

\bibitem Borges, J. L. (1946).  Del rigor en la ciencia.  {\em Los Anales de Buenos Aires} {\bf 1} (3).
  
\bibitem Cleveland, W. S. (1985).  {\em The Elements of Graphing Data}. Monterey, Calif.: Wadsworth.

\bibitem Cui Q., Ward M., Rundensteiner E., and Yang J. (2006). Measuring data abstraction quality in multiresolution visualizations. {\em IEEE Transactions on Visualization and Computer Graphics} {\bf 12} (5), 709--716.

\bibitem Fabrikant, S. I. (2004). Abstraction and scale in spatialisation. In {\em Proceedings of the National Carographic Conference},  35--43. New Zealand Cartographic Society.
  
\bibitem Feller, A., Gelman, A., and Shor, B. (2012).  Red state / blue state divisions in the 2012 presidential election.  {\em The Forum} {\bf 10}, 127--141.

\bibitem Gelman, A., and Azari, J. (2017).  19 things we learned from the 2016 election (with discussion). {\em Statistics and Public Policy} {\bf 4} (1), 1--10. 

\bibitem Gelman, A., Park, D., Shor, B., and Cortina, J. (2009). {\em Red State, Blue State, Rich State, Poor State: Why Americans Vote the Way They Do}, second edition. Princeton University Press.

\bibitem Gelman, A., Vehtari, A., Simpson, D., Margossian, C. C., Carpenter, B., Yao, Y., Bürkner, P. C., Kennedy, L., Gabry, J., and Modrák, M. (2020).  Bayesian workflow. \url{https://arxiv.org/abs/2011.01808}

\bibitem Guthrie, A. (1972). Modern semantics can help accounting. {\em Journal of Accountancy} {\bf 133}, 56--63.

\bibitem Hayakawa, S. I. (1939).  {\em Language in Action}. New York: Harcourt, Brace.

\bibitem Hurst, J. W., and Walker, H. K., eds. (1972). {\em The Problem-Oriented System}. New York: Medcom.

\bibitem Inselberg, A. (1985).  The plane with parallel coordinates.  {\em The Visual Computer} {\bf 1}, 69--91.

\bibitem Jahnke, M., Krisp, J. M., and Kumke, H. (2011). How many 3D city models are there?---A typological try. {\em Cartographic Journal} {\bf 48}, 124--130.

\bibitem Korzybski, A. (1933).  {\em Science and Sanity}, 747--761.  New York:  International Non-Aristotelian Library Publishing Company.

\bibitem Leviton, L. C. (2015). Evaluation practice and theory: Up and down the ladder of abstraction. {\em American Journal of Evaluation} {\bf 36}, 238--242.

\bibitem Metzgar, C. (1978). Hayakawa's cow \& safety. {\em Professional Safety} {\bf 23} (8), 44--48.

\bibitem Moffett, J. (1968). Teaching the universe of discourse.  Boston:  Houghton Mifflin.  Available at  \url{https://wac.colostate.edu/books/landmarks/moffett/universe/}.

\bibitem Panckoucke, C. J. (1783). {\em Encyclopédie Méthodique Marine}. Paris: Chez Panckoucke.

\bibitem Tufte, E. R. (1983).  {\em The Visual Display of Quantitative Information}. Cheshire, Conn.: Graphics Press.

\bibitem Viola, I., and Isenberg, T. (2018).  Pondering the concept of abstraction in (illustrative) visualization.  {\em IEEE Transactions on Visualization and Computer Graphics} {\bf 24} (9), 2573--2588.
  
\bibitem Wainer, H., and Friendly, M. (2021).  Displaying information:  From woolly mammoths to the Great Migration.  In {\em International Encyclopedia of Education}, 4th edition, ed.\ R. Tierney, F. Rizvi, K. Ercikan, and G. Smith. Philadelphia: Elsevier.
  
\bibitem Wickham, H. (2016).  {\em ggplot2: Elegant Graphics for Data Analysis}. New York: Springer-Verlag. \url{ggplot2.tidyverse.org}

\bibitem Wilkinson, L. (2005).  {\em The Grammar of Graphics}, second edition. New York: Springer-Verlag.

\end{document}